\documentclass[aps,amsmath,amssymb,reprint,superscriptaddress]{revtex4-2}

\usepackage{graphicx}%
\usepackage{dcolumn}% Align table columns on decimal point
\usepackage{bm}% bold math
\usepackage[colorlinks,linkcolor=blue,urlcolor=black,citecolor=blue]{hyperref}%
\usepackage[caption=false]{subfig}
\newcommand{\phantomsubfloat}[1]{
    {% apply caption setup only temporarily
        \captionsetup[subfloat]{farskip=0pt,captionskip=0pt}
        \captionsetup[subfigure]{labelformat=empty}
        \subfloat{#1}
    }%
}
\usepackage[compress]{cleveref}
\crefname{equation}{Eq.}{Eqs.}
\Crefname{equation}{Equation}{Equation}
\crefname{figure}{Fig.}{Fig.}
\Crefname{figure}{Figure}{Figure}
\crefname{subfigure}{Fig.}{Fig.}
\Crefname{subfigure}{Figure}{Figure}
\crefrangelabelformat{subfigure}{#3{#1}#4--#5{#2}#6}
\creflabelformat{equation}{#2\textup{#1}#3}
\usepackage{siunitx}

\newcommand{\SIO}{Sr$_2$IrO$_4$}
\newcommand{\CROb}{Ca$_3$Ru$_2$O$_7$}
\newcommand{\CRTO}{Ca$_3$(Ru$_{0.99}$Ti$_{0.01}$)$_2$O$_7$}
\newcommand{\CRTOx}{Ca$_3$(Ru$_{1-x}$Ti$_x$)$_2$O$_7$}

\DeclareUnicodeCharacter{2212}{-}

\begin{document}
\title{Magnetic field-driven phase switching in the antiferromagnetic\\ Mott insulator \CRTO}
%\title{Magnetic field-driven phase switching in the antiferromagnetic and strongly correlated bilayer oxide \CRTO}
\author{Ksenia~S.~Rabinovich}
%\email{K.Rabinovich@fkf.mpg.de}
\thanks{These authors contributed equally to this work.}
\author{Tim~Priessnitz}
%\email{T.Priessnitz@fkf.mpg.de}
\thanks{These authors contributed equally to this work.}
\author{Nils~Gross}
\thanks{These authors contributed equally to this work.}
\affiliation{
Max Planck Institute for Solid State Research, Heisenbergstra{\ss}e~1,
70569 Stuttgart, Germany}
\author{George~Jackeli}
\affiliation{
Max Planck Institute for Solid State Research, Heisenbergstra{\ss}e~1,
70569 Stuttgart, Germany}
\affiliation{
Andronikashvili Institute of Physics, Tbilisi State University, Tamarashvili str. 6, 0177 Tbilisi, Georgia}
\author{Maximilian~J.~Krautloher}
\author{Pascal~Reiss}
\author{Eberhard~J.~Goering}
\author{Jurgen~H.~Smet}
\author{Bernhard~Keimer}
\author{Alexander~V.~Boris} 
\email{A.Boris@fkf.mpg.de}
\affiliation{
Max Planck Institute for Solid State Research, Heisenbergstra{\ss}e~1,
70569 Stuttgart, Germany}

\date{\today}% It is always \today, today,

\begin{abstract}
A bandwidth-controlled antiferromagnetic Mott-insulating phase in \CRTOx \ is realized through isovalent substitution at the Ru site. For a dilute  substitution with only 1\% Ti, the Mott insulator ground state remains nearly degenerate with the ground state of pristine \CROb, where the Ru moments are ferromagnetically aligned within the metallic RuO$_2$ bilayers stacked in an antiferromagnetic fashion. The exceptionally shallow free energy landscape of this doped compound arises from intertwined electron–electron and electron–lattice interactions. This makes its magnetic and transport properties highly sensitive to external perturbations. We systematically investigated magnetic-field-induced phase switching in \CRTO \ to explore its magnetic $H$-$T$ phase diagram. 
With the field applied along the easy $b$-axis, parallel to the antiferromagnetic moments, the magnetization exhibits a first-order spin-flop transition at $\approx\SI{6}{\tesla}$, indicating reorientation of the $\rm Ru$ moments perpendicular to the field. The transition is accompanied by a decrease in the electrical resistance, but the spin-flop phase remains insulating. Above \SI{10.5}{\tesla}, all Ru moments align with the $b$-axis, resulting in a forced ferromagnetic metallic phase. In contrast, neither spin-flop nor forced-ferromagnetic phases are observed up to \SI{14}{\tesla}, when the field is applied along the $a$-axis. While the electronic kinetic energy and the electron-lattice coupling contribute to the free-energy balance of this system, the resulting $H$-$T$ phase diagram is remarkably simple and closely resembles that of a canonical anisotropic antiferromagnet, albeit with substantially renormalized critical fields.

\end{abstract}

\maketitle

\section{Introduction}
In $4d$ transition-metal layered perovskites, in particular the ruthenates, the energy scales of competing collective instabilities, intra-atomic spin-orbit coupling, and electronic correlation strength are comparable. As a consequence, they host a variety of nearly degenerate ground states from superconductivity over Mott insulators to altermagnetism   \cite{Cao1999,Cao1997,Ikeda2000,Maeno2001,cuono2023orbital,leon2025hybrid}. The degeneracy implies that even minor chemical substitution or slight external stimuli – such as pressure, electric field, and illumination – can significantly affect the phase behavior and macroscopic properties \cite{Snow2002,Bertinshaw2019,Zhang2019,Suen2024,Rabinovich2024}. A notable example of such electronic phase sensitivity to small perturbations is bilayer perovskite calcium ruthenate \CROb. Its ground state is a correlated polar metal with itinerant electrons within the antiferromagnetically (AFM) stacked ferromagnetic (FM) RuO$_2$ bilayers \cite{Cao1997,Bao2008,Bohnenbuck2008,Sokolov2019,Markovic2020,Faure2023}. Remarkably, substituting just 1\% of $\rm Ru$ with isovalent $\rm Ti$, and thus minimally tuning the effective Ru electronic bandwidth, transforms \CROb \ into an AFM Mott insulator \cite{Ke2011,Tsuda2013,Krautloher2018,Rabinovich2024}, placing it in a regime where competing FM metallic and AFM insulating phases coexist at nearly identical energies.

At room temperature, both \CROb \ and \CRTO \ are paramagnetic metals (PM-M) with an orthorhombic crystal structure, characterized by the polar space group $Bb2_1m$. Upon cooling, pristine \CROb \ undergoes two consecutive isostructural magnetic transitions: an A-type AFM-$a$ ordering at $T_\mathrm{N} = \SI{56}{\kelvin}$, with FM alignment along the $a$-axis within the RuO$_2$ bilayers, followed by a metamagnetic spin reorientation transition to the AFM-$b$ phase at $T_\mathrm{s} = \SI{48}{\kelvin}$, stabilizing a polar metallic state with FM moments aligned along the $b$-axis. In contrast, the 1\% $\rm Ti$-substituted compound \CRTO \ undergoes a first-order metal-to-insulator transition (MIT) concomitant with G-type AFM ordering, i.e. AFM alignment of the $\rm Ru$ magnetic moments both within and between the bilayers along the $b$-axis (G-AFM-$b$), at $T_\mathrm{MIT}=T_\mathrm{N}=\SI{55}{\kelvin}$. This transition is accompanied by significant disproportionate changes in the crystal lattice parameters and the itinerant electronic behavior is replaced by charge localization coupled in a robust manner to the spin alignment.

The confluence of electron–electron and electron-lattice interactions renders the material extraordinarily sensitive to external perturbations such as light illumination \cite{Rabinovich2024}. An exceptionally shallow free-energy landscape of its G-AFM-$b$ Mott insulator ground state, nearly degenerate to the metallic AFM-$a,b$ state, highlights \CRTO \ as a unique model system for building and testing a theory of Mott transition dynamics in the presence of strong electron-lattice coupling. Although the response to magnetic fields has been investigated extensively in \CRTOx, reported $H$-$T$ phase diagrams vary wildly in complexity depending on the level of Ti substitution and, presumably, preparation details. Disentangling the roles of spin reorientation, lattice distortions, and electronic localization to the evolution and development of the various phases has proven challenging. Exploring the transport and magnetic properties of the $x=1\%$ compound across an extended range of magnetic fields is therefore highly desirable to clarify the underlying physics.

From a technological perspective, the exceptional light sensitivity of the MIT suggests potential applications in designing nanodevices with quantum-level photosensitivity \cite{Rabinovich2024}. Recent studies have also demonstrated that the MIT in \CRTO \ at heterointerfaces with the relativistic $5d$ Mott insulator \SIO \ offers an effective method for tuning magnons, indicating a promising avenue for AFM magnonics \cite{Shrestha2025}. Furthermore, AFM insulators with a preferred magnetization direction, such as \CRTO, are attractive alternatives to ferromagnets for spintronic applications, since they possess a field-switchable spin-flop transition that combines ultrafast dynamics with low magnetic susceptibility \cite{Baltz2018,Gomonay2018,Antropov2021,Chen2024}.

In \CRTO, the magnetic anisotropy is intrinsically linked to its electronic configuration: the $\rm Ru^{4+}$ ions adopt a low-spin $S=1$ configuration, where the extended $4d$ orbitals, associated with a strong ligand field, stabilize a state dominated by spin-orbit coupling and crystal field effects. This pronounced anisotropy turns \CRTO \ into an ideal platform for field-controlled spin reorientation. While conventional anisotropic AFMs exhibit spin-flop transitions determined by a simple balance between exchange and anisotropy energies, in strongly correlated systems with magnetoelastic coupling, the physics is far richer, as spin reorientation, lattice distortion, and electronic localization are intertwined. This can potentially lead to a significant renormalization of the critical switching fields. This phenomenon is addressed in this work by systematically recording the $H$-$T$ phase diagram of \CRTO \ under fields up to \SI{14}{\tesla} applied along the in-plane crystallographic axes $a$ and $b$, i.e., the magnetic hard and easy directions, respectively, and by comparing the magnetization and magnetic susceptibility behavior with transport data.

\begin{figure}[h!]
\includegraphics[width=8cm]{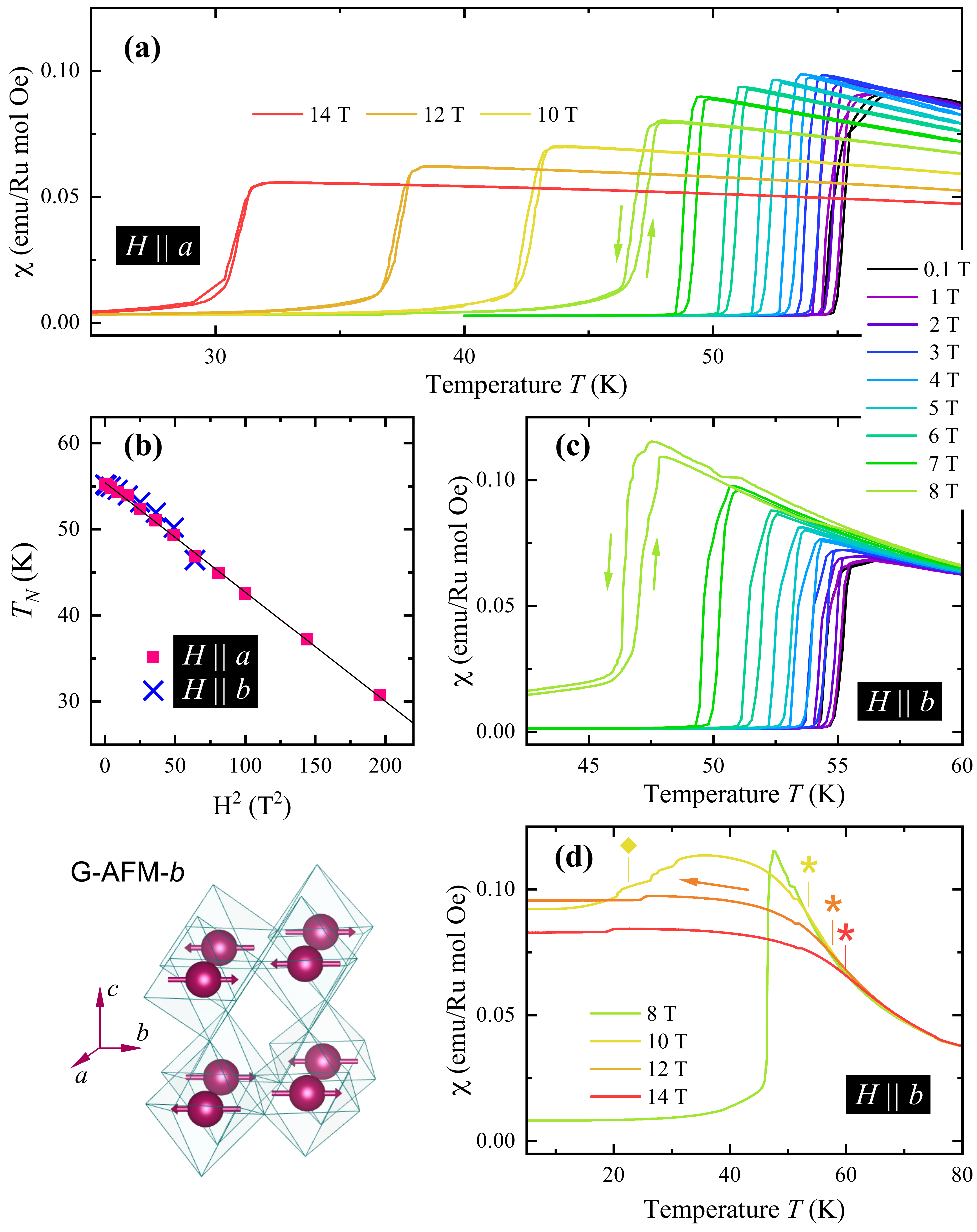}
\phantomsubfloat{\label{fig:figure01_a}}
\phantomsubfloat{\label{fig:figure01_b}}
\phantomsubfloat{\label{fig:figure01_c}}
\phantomsubfloat{\label{fig:figure01_d}}
\caption{\label{fig:1} Temperature dependence of magnetic susceptibility $\chi (T)$ of \CRTO \ measured at different magnetic fields: (a) $\chi (T)$ measured at magnetic fields up to 14 T applied along the $a$-axis. (b) The G-AFM to PM phase transition temperature from field cooled (FC) $\chi (T)$ measurements plotted against magnetic field squared, with $H||a$-axis (red squares) and $H||b$-axis (blue crosses). (c,d) $\chi (T)$ measured with $H||b$-axis at field below (c) and above (d) 8T. In (d), the diamond and stars correspond to the positions of the broad maximum and minima in the derivatives of FC $\chi (T)$, respectively. The image of Ru moments (red) surrounded by oxygen octahedra illustrates the magnetically ordered G-AFM-$b$ state at zero field.
}
\end{figure}

\section{Experimental methods}

High-quality single-domain crystals of \CRTO \ were grown using an optical floating-zone technique and selected based on the reflectivity contrast in polarized-light optical microscopy. Careful sample selection avoided artifacts from multiphase behavior and phase separation that have been reported to affect the magnetic and transport properties of \CRTOx \ at higher $\rm Ti$ substitution levels \cite{Peng2016a,Peng2016b,Zhu2016}. High-resolution X-ray diffraction (XRD) and THz spectroscopy confirmed their phase purity and monodomain character \cite{Krautloher2018,Rabinovich2024}. Typical crystal dimensions were $2\times1\times0.2$ mm$^3$. Magnetic phase transitions in \CRTO \ were identified from anomalies in the temperature-dependent magnetic susceptibility $\chi(T)$ under various magnetic fields and/or in the isothermal magnetization $M(H)$ below $T_\mathrm{N}=\SI{55}{\kelvin}$. Magnetization was measured up to \SI{7}{\tesla} with a Quantum Design SQUID magnetometer and up to \SI{14}{\tesla} with a Quantum Design Physical Properties Measurement System (PPMS) equipped with a Vibrating Sample Magnetometer (VSM) option. Similarly, electronic transitions were detected from anomalies in the resistivity recorded as functions of temperature and magnetic field. The in-plane resistivity was measured up to \SI{14}{\tesla} and down to \SI{5}{\kelvin} on oriented samples contacted with conductive silver glue. Through sample rotation the magnetic field was aligned with the crystal axes.  To address uncertainties in absolute resistivity arising from anisotropy and possible current inhomogeneity, we employed both two- and four-terminal configurations. Unless stated otherwise, we report the normalized resistivity change, $(\rho-\rho_0)/\rho_\mathrm{max}$ , where $\rho_0$ and $\rho_\mathrm{max}$ are the zero field and maximum resistivity, respectively. Low-frequency lock-in detection at 38 Hz was used to improve the signal-to-noise ratio and served as a proxy for the DC resistivity. To avoid current-induced phase transformations known from the monolayer member of the same Ruddlesden–Popper series \cite{nakamura2013electric,okazaki2013current,Zhang2019}, an AC excitation current as low as \SI{200}{\nano\ampere} was used. Samples were appropriately mounted to minimize strain and to avoid affecting the structural transition.

\begin{figure}[b!]
\includegraphics[width=\linewidth]{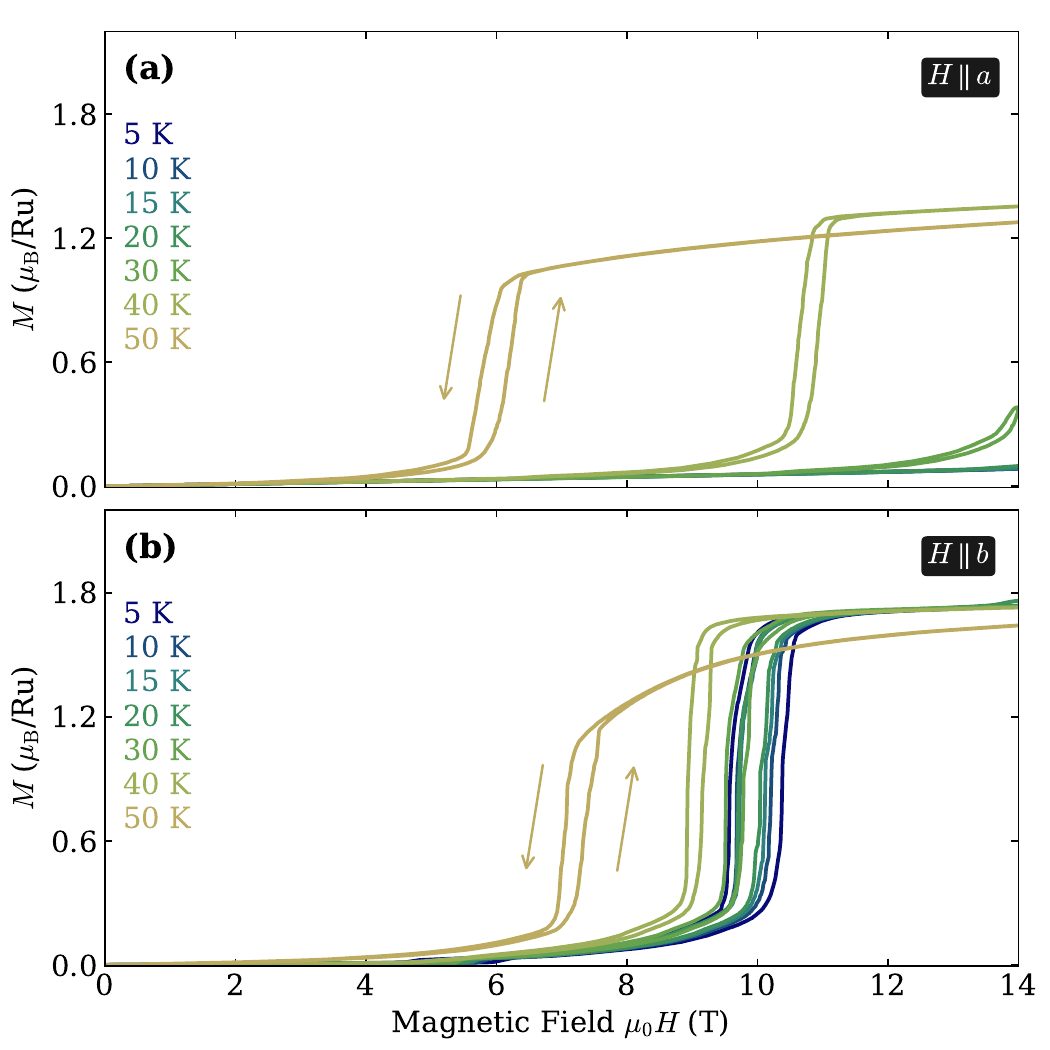}
\phantomsubfloat{\label{fig:figure02_a}}
\phantomsubfloat{\label{fig:figure02_b}}
\caption{\label{fig:2}
(a,b) Magnetization of \CRTO \ as a function of the magnetic field, measured at temperatures below $T_\mathrm{N}$. Magnetic fields up to \SI{14}{\tesla} are applied along the $a$-axis (a) and $b$-axis (b). In (b), the magnetic moment at \SI{14}{\tesla} for temperatures below \SI{40}{\kelvin} reaches about \SI{1.73}{\mu_B\per Ru}, indicating a trend towards saturation.
}
\end{figure}

\section{Results and discussion}
\subsection{Magnetic properties}
 The magnetic susceptibility $\chi(T)$ of \CRTO \ exhibits a sharp drop with decreasing temperature below $T_\mathrm{N}$, accompanied by pronounced thermal hysteresis. Figure \ref{fig:1} illustrates that the hysteresis loop shifts to lower temperatures with increasing applied magnetic field. When the magnetic field is aligned along the $a$-axis (\cref{fig:figure01_a}), the transition temperature decreases quadratically with increasing field up to \SI{14}{\tesla}. A linear fit of $T_\mathrm{N}$ versus $H^2$, shown in \cref{fig:figure01_b}, yields 
\begin{equation}
    \label{eq:TNvsH2}
    T_\mathrm{N}(H)=T_\mathrm{N}(0)-\alpha H^2,
\end{equation}
with $T_\mathrm{N}(0) = 55.4\pm\SI{0.1}{\kelvin}$ and $\alpha= 0.127\pm\SI{0.004}{\kelvin\per\tesla^2}$. This quadratic behavior reflects the competition between the Zeeman energy  and exchange interactions in a collinear AFM, leading to a characteristic field-dependent suppression of the Néel temperature. Within a molecular-field approximation, the quadratic coefficient $\alpha$ is given by \cite{Heller1966,Shapira1970}:
\begin{equation}
\alpha = \frac{3(2S^{2} + 2S + 1)}{40(S + 1)^{2}} 
\left[ \frac{\chi_\mathrm{N}}{M_\mathrm{s}} \right]^{2} T_\mathrm{N}(0),  
\end{equation}
where $S$ is the spin quantum number, $\chi_\mathrm{N}$ is the susceptibility at $T_\mathrm{N}(0)$, and $M_\mathrm{s}$ is the saturation sublattice magnetization. Using $S=1$, $\chi_\mathrm{N}=0.065\;\rm emu/Ru \ mol \ Oe$ from \cref{fig:figure01_a}, $M_\mathrm{s}= \SI{0.75}{\mu_B\per Ru}$ \cite{Krautloher2018,BertinshawNSDraft2020} and  $T_\mathrm{N}=\SI{55.4}{\kelvin}$, we obtain $\alpha\approx0.13$ K/T$^2$. This is consistent with the value extracted from the experiment. By extrapolating $T_\mathrm{N}(H^2)$ down to zero temperature, we find that the complete suppression of the AFM order requires a magnetic field of $\approx\SI{20.9}{\tesla}$. Below $T_\mathrm{N}$, both zero-field-cooled (ZFC) and field-cooled (FC) $\chi(T)$ curves (\cref{fig:figure01_a}) drop to a common minimum value, indicative of AFM ordering. This behavior is maintained up to \SI{14}{\tesla} for magnetic fields aligned along the $a$-axis, without the emergence of new magnetic phases.

For fields applied along the $b$-axis, the same behavior persists up to \SI{7}{\tesla}, as shown in \cref{fig:figure01_c} and denoted by cross symbols in \cref{fig:figure01_b}. At \SI{8}{\tesla}, however, the $\chi(T)$ develops a finite background below $T_\mathrm{N}$ accompanied by an increasing disparity between the FC and ZFC measurements. The enhanced low-temperature $\chi(T)$ indicates the presence of uncompensated magnetic moments. Applying such a strong field competes with the AFM exchange interaction, promoting spin alignment along the field direction. As the magnetic field is increased further, the susceptibility cusp becomes progressively smeared out. \Cref{fig:figure01_d} thus suggests that the collapse of the AFM order leads to a ferromagnetic-like state at fields above \SI{10}{\tesla}. This is indicative of the presence of a net magnetic moment, which will be discussed in the context of magnetization measurements in the following section. The crossover point, where thermal fluctuations get strong enough to cause a reduction of said net moment, is determined by extrema in the derivatives of the FC curves. These extrema are marked by star symbols in \cref{fig:figure01_d}.

To unambiguously identify the nature of the observed magnetic ground states, we now examine the field-dependent magnetization $M(H)$. \Cref{fig:figure02_a} shows that at fields along the $a$-axis and at low temperatures, $M(H)$ remains linear in the field up to \SI{14}{\tesla}. However, as the temperature approaches $T_\mathrm{N}(0)$, a field-driven first-order transition emerges at $\approx \SI{12}{\tesla}$ and $\approx \SI{6}{\tesla}$ at \SI{40}{\kelvin} and \SI{50}{\kelvin}, respectively. After a steep increase of about 1 $ \mu_B / \rm Ru$, $M(H)$ does not saturate but continues to rise in a nearly linear fashion. The critical magnetic fields observed during these field sweeps are in good agreement with the G-AFM to PM phase boundary described by the quadratic relationship (\cref{eq:TNvsH2}) that was derived for $\chi(T)$ in \cref{fig:figure01_b}. So far, this behavior is fully consistent with the mean-field nature of the phase transition, indicating a conventional quantum critical point at a magnetic field of $\approx\SI{20.9}{\tesla}$.

Conversely, at low temperatures, when the field is increased along the $b$-axis, $M(H)$ exhibits a sharp jump with a clear trend towards saturation at high fields. \Cref{fig:figure02_b} shows that for temperatures below \SI{40}{\kelvin}, the (nearly) saturated magnetic moment corresponds to 1.73 $\mu_B$ per Ru at \SI{14}{\tesla}, which is close to the Ru magnetic moment in pristine \CROb \ \cite{Cao1997,Bao2008,Faure2023}. Hence, we conclude that above the corresponding critical fields, $H_b^\mathrm{sat}$, the G-AFM phase collapses into a forced FM state, i.e. the fully saturated state of the PM phase. Notably, at \SI{50}{\kelvin}, close to $T_\mathrm{N}(0)$, $M(H)$ does not saturate above the critical field of $\approx \SI{7}{\tesla}$, but closely resembles the behavior for fields along the $a$-axis. Accordingly, this transition corresponds to the $H$-$T$ phase line represented by the cross symbols in \cref{fig:figure01_b} along the quadratic field dependence of $T_\mathrm{N}(H)$. We note that a similar field-induced transition was reported for $x=3\%$ $\rm Ti$-substituted \CRTOx \ at fields below $H_b^\mathrm{sat}$ \cite{Zhu2016}, where incomplete saturation of the magnetic moment was attributed to a canted AFM structure comprising an AFM component along the $a$-axis and a FM component along the $b$-axis. However, recent neutron diffraction measurements on the one \CRTO \ single crystal from the same batch as used in the present study confirm nearly complete FM alignment of the $\rm Ru$ moments in the forced FM state above $H_b^\mathrm{sat}$ \cite{BertinshawNSDraft2020}.

\begin{figure}
\includegraphics[width=\linewidth]{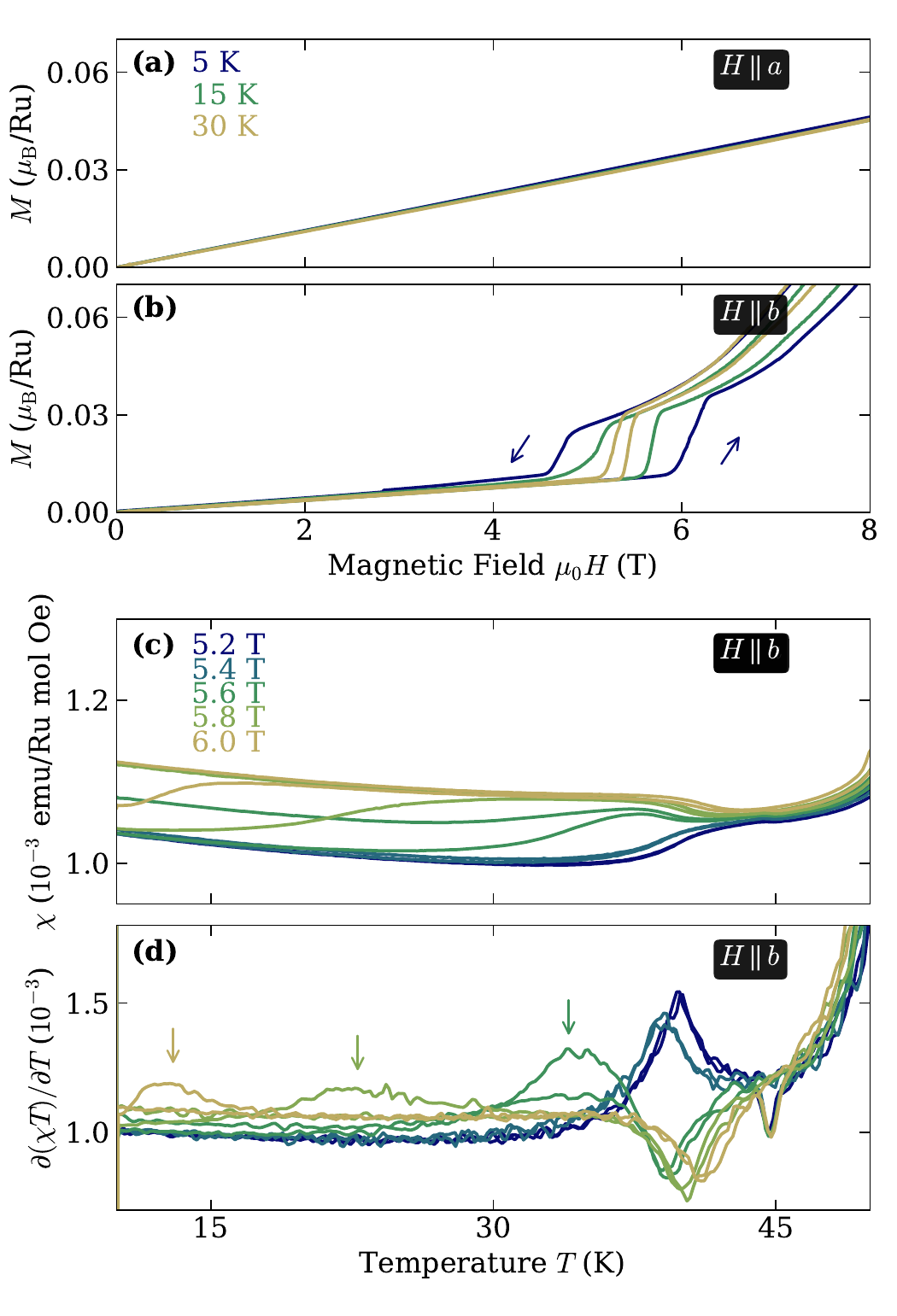}
\phantomsubfloat{\label{fig:figure03_a}}
\phantomsubfloat{\label{fig:figure03_b}}
\phantomsubfloat{\label{fig:figure03_c}}
\phantomsubfloat{\label{fig:figure03_d}}
\caption{\label{fig:3} 
(a,b) Field-dependent magnetization of \CRTO \ measured at low temperatures with fields applied along the $a$-axis (a) and $b$-axis (b). In (b), $M(H)$ measured along the easy $b$-axis exhibits a sharp, step-like hysteretic change at the spin-flop transition field, $H_b^\mathrm{f}$, indicating the reorientation of the Ru moments from AFM order into a canted spin-flop state, perpendicular to the field. (c) Magnetic susceptibility $\chi (T)$ measured at magnetic fields near $H_b^\mathrm{f}$ applied along the $b$-axis. The zero-field-cooled (ZFC) and field-cooled (FC) $\chi (T)$ curves diverge below the critical temperatures corresponding to the applied fields, as determined by the peaks in the derivatives of ZFC $\chi (T)$, indicated by arrows in (d).
}
\end{figure}

\begin{figure}[t!]
\includegraphics[width=7.5cm]{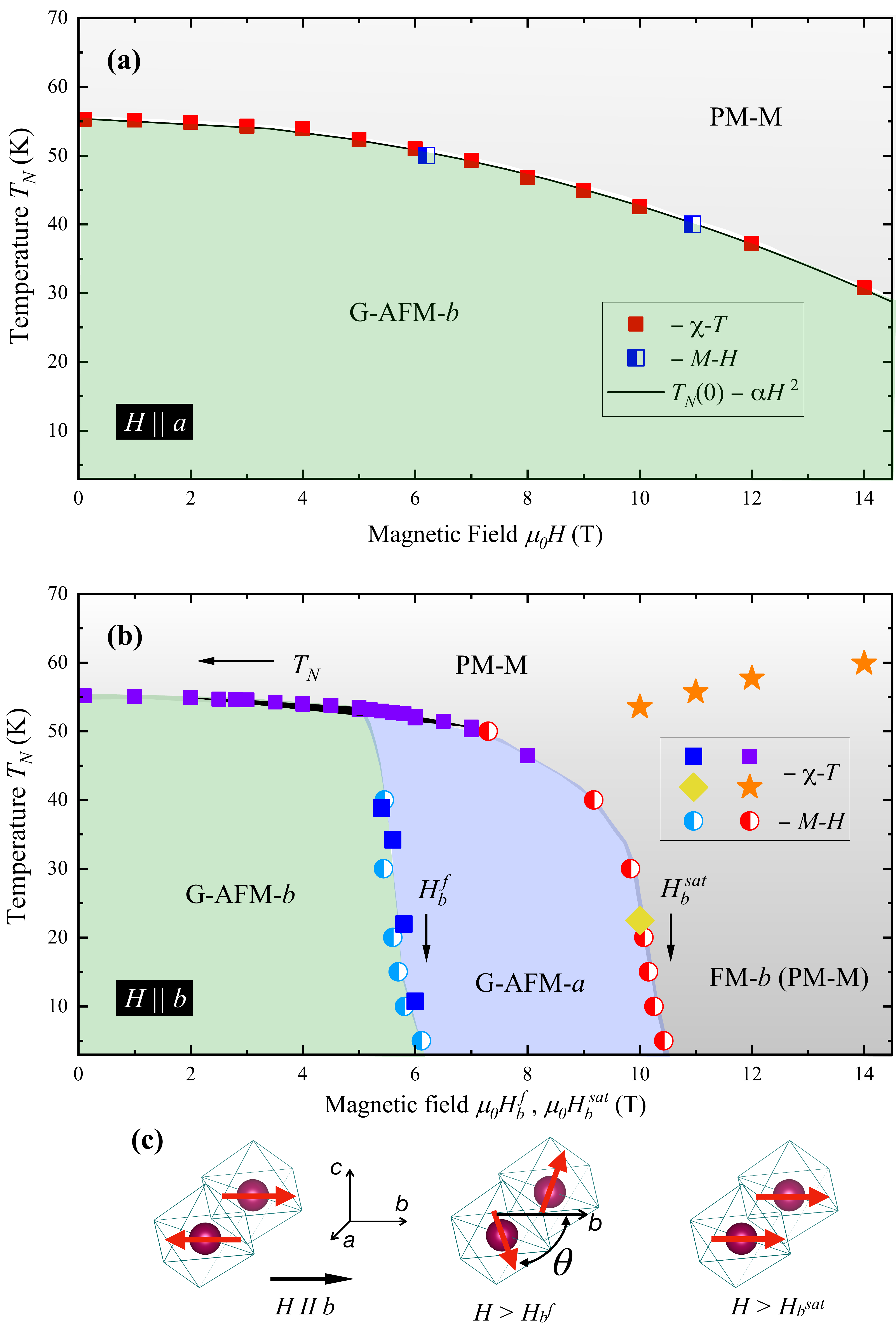}
\phantomsubfloat{\label{fig:figure04_a}}
\phantomsubfloat{\label{fig:figure04_b}}
\phantomsubfloat{\label{fig:figure04_c}}
\caption{\label{fig:4}
(a,b) Magnetic temperature–field ($T$-$H$) phase diagrams of \CRTO \ with the magnetic field applied along the $a$-axis (a) and $b$-axis (b). The phase diagrams are derived from critical temperatures observed in temperature-dependent magnetic susceptibility ($\chi$-$T$) as shown in \cref{fig:1,fig:figure03_c,fig:figure03_d} (filled symbols), and critical fields from the field-dependent magnetization ($M$-$H$) as shown in \cref{fig:2,fig:figure03_a,fig:figure03_b} (half-filled symbols). In (a), the critical $T_N$ temperature for AFM ordering  decreases quadratically with increasing field along the $a$-axis. The critical field values correspond to temperature-driven transitions from the G-AFM-$b$ to the paramagnetic PM-M state, indicating that the magnetic field along the $a$-axis does not induce a new magnetic order in \CRTO. (b) When the magnetic field is applied along the easy $b$-axis, the low-temperature magnetization exhibits a first-order spin-flop transition at $H_b^\mathrm{f}$, indicating the reorientation of Ru moments perpendicular to the field. Increasing the field above $H_b^\mathrm{sat}$ aligns all Ru moments with the $b$-axis, resulting in a forced FM phase. With increasing $T$ the forced FM state is suppressed by thermal fluctuations restoring PM behavior above the crossover temperatures depicted by star symbols, which correspond to the broad minima in the derivatives of FC $\chi (T)$ marked by stars in \cref{fig:figure01_d}. The temperature variation of $H_b^{f}$ and $H_b^{sat}$ forms a typical phase diagram for an anisotropic AFM with the field parallel to the magnetization-easy axis. (c) Schematic illustration of the neighboring Ru magnetic moment alignment corresponds to the G-AFM-$b$ ground state, spin-flop G-AFM-$a$ phase, and field-forced FM-$b$ phase, from left to right respectively.
}
\end{figure}

In a conventional AFM with uniaxial anisotropy, upon application of a magnetic field along the easy axis, i.e. parallel to the sublattice magnetic moments, the field-driven spin-flip transition from an AFM to forced FM ground state is typically preceded by a spin-flop transition. This spin-flop transition is characterized by a sudden reorientation of the AFM sublattices perpendicular to the applied field, enabling a progressive canting towards full alignment as the field is increased further. To search for this intermediate spin-flop phase as the magnetic field increases along the easy $b$-axis, before transitioning into the high-field forced FM state, we take a closer look on the low temperature $M(H)$ for fields below \SI{8}{\tesla}. First, \cref{fig:figure03_a} shows that when the field is applied along the hard $a$-axis, perpendicular to the AFM moments, it causes the Ru moments to cant in the field's direction. This results in a linear dependence of the magnetization, expressed as
\begin{equation}\label{eq:linearM}
    M_i(H)=\beta_iH,
\end{equation}
where the index $i$ refers to the crystallographic direction, here the magnetic hard $a$-axis and $\beta_a=(5.75\pm0.01)\times10^{-3}\rm \mu_B/Ru\ T^{-1}$ at $T=\SI{5}{\kelvin}$.

In contrast, for the field along the easy $b$-axis (parallel to the AFM moments), the net magnetization at first increases linearly but at a slow pace, with $\beta_b=(1.91\pm0.02)\times10^{-3}\rm \mu_B/Ru\ T^{-1}$ at $T=$ 5 K,  consistent with partial flips of the Ru moments that were initially antiparallel to the field. However, as the field is ramped up further, a hysteretic jump in the magnetization $\Delta M_b^\mathrm{f}$ occurs above the critical field $H_b^\mathrm{f}$, as shown in \cref{fig:figure03_b}. At $T=\SI{5}{\kelvin}$ and $H_b^\mathrm{f}\approx\SI{6}{\tesla}$, $\Delta M_b^\mathrm{f}\approx 0.025\ \rm \mu_B/Ru$. Notably, after the abrupt increase, the magnetization value coincides with that obtained when the same field is applied along the $a$-axis
\begin{equation}
    M_b(H)\approx M_a(H)+\Delta M^\mathrm{f}_b    
\end{equation}
at $H\gtrapprox H_b^\mathrm{f}$, indicating a field-induced reorientation of the spin structure into a canted arrangement perpendicular to the field. This observed spin-flop transition is first-order and accompanied by a hysteresis that narrows progressively with increasing temperature as a result of thermal fluctuations. The phase boundary between the G-AFM-$b$ ground and G-AFM-$a$ spin-flop states can also be identified from the temperature dependencies of $\chi (T)$ measured at magnetic fields applied along the $b$-axis near but below $H_b^\mathrm{f}$. \Cref{fig:figure03_c} shows that the ZFC and FC $\chi (T)$ curves diverge below the critical temperatures corresponding to the applied fields. The phase boundary is then defined by the corresponding peak positions in the temperature-weighted derivatives of the ZFC $\chi (T)$, indicated by arrows in \cref{fig:figure03_d}. We point out, that in the case of a conventional anisotropic AFM, the specific heat is directly related to  $\partial (T\chi)/ \partial T$ up to a proportionality constant that only slowly varies with $T$ \cite{fisher1962relation}.

Having identified all magnetic phases from the observed anomalies in $\chi(T)$ and $M(H)$ measurements, we can now compile the $H$-$T$ phase diagram of \CRTO. For the external field $H$ applied parallel to the $a$-axis, perpendicular to the low-$T$ AFM moments, the phase diagram in \cref{fig:figure04_a} contains only two phases: an antiferromagnetic (G-AFM-$b$) and a paramagnetic (PM). When $H$ is applied along the easy $b$-axis, the phase transitions as a function of $T$ ($T_\mathrm{N}$) and $H$ ($H_b^\mathrm{f}$,$H_b^\mathrm{sat}$) produce three distinct regions in \cref{fig:figure04_b} separated by phase boundaries: the G-AFM-$b$ phase, the spin-flop G-AFM-$a$ phase, and the forced ferromagnetic FM-$b$ phase that evolves into the PM state at elevated temperatures. At zero field in the G-AFM-$b$ state, magnetizations of the two equivalent $\rm Ru$-moment sublattices are directed along the $b$-axis, equal in magnitude and antiparallel. Above $H_b^\mathrm{f}$, in the spin-flop G-AFM-$a$ phase, the sublattice moments remain equal and nearly antiparallel but rotate to be nearly perpendicular to $H\parallel b$, forming equal, nonzero angles with $H$ - an arrangement analogous to G-AFM-$b$ for $H\parallel a$ in \cref{fig:figure04_a}. The star symbols at high fields, derived from the $\chi(T)$ data in \cref{fig:figure01_d}, mark the temperatures above which thermal fluctuations restore PM behavior in the forced FM$-b$ state.

Magnetic phase boundaries in anisotropic Heisenberg AFMs are typically determined by minimizing the total energy, which reflects the competition among the AFM exchange, anisotropy, and Zeeman terms. Inspired by the experimentally determined phase diagram, reminiscent of that of a conventional anisotropic AFM, we consider a simple spin Hamiltonian, taking into account the bilayer nature of our system with isotropic Heisenberg inter-site exchange and uniaxial single-ion anisotropy in a fiel:
\begin{multline}\label{eq:Hamiltonian}
    \mathcal{H} = J_{\parallel} \sum_{\langle i j \rangle} \left[ \vec{S}_{1i} \cdot \vec{S}_{1j} + \vec{S}_{2i} \cdot \vec{S}_{2j} \right] +J_{\bot} \sum_{\langle i \rangle} \vec{S}_{1i} \cdot \vec{S}_{2i}\\ 
    - D \sum_{i} \left[ (S_{1i}^{b})^{2} + (S_{2i}^{b})^{2} \right]\\ 
    - \sum_{i} g \mu_{B} \left[ \vec{S}_{1i} + \vec{S}_{2i} \right] \cdot \vec{H},
\end{multline}
where  $\vec{S}_{1i,2i}$ and $\vec{S}_{1j,2j}$ are the spin operators for neighboring sites $i$ and $j$ in layers $1$ and $2$ of the bilayer, $J_{\parallel}$ and $J_{\bot}$ are positive in-plane and inter-layer exchange coupling constants, respectively, and $D$ is the orthorhombic single-ion anisotropy term for $b$-axis distortions to the local cubic crystal symmetry. When the external magnetic field is aligned with the hard $a$-axis,  the total magnetic energy per spin is
\begin{multline}
    E(\theta) = -(2J_{\parallel} +\frac{1}{2}J_{\bot})S^2\cos{2\theta}
    -DS^2\cos^2{\theta}\\
    -g\mu_BH_aS\sin{\theta}.
\end{multline}
Here, $\theta$ is the symmetric canting angle of the Ru sublattice moments away from the easy $b$-axis toward $\vec{H}\parallel a$. As $H_a$ increases, this canting yields a gradual increase of the net magnetization described by \cref{eq:linearM} with $\beta_a=g \mu_b S/H_a^\mathrm{sat}$ up to saturation at 
\begin{equation}
    g\mu_bH_a^\mathrm{sat}=2S(4J_{\parallel}+J_{\bot}+D).
\end{equation}
When the field is applied along the easy $b$-axis, the two critical fields are determined by 
\begin{equation}
    g\mu_bH_b^\mathrm{f}=2S\sqrt{D(4J_{\parallel}+J_{\bot}-D)}
\end{equation}
and 
\begin{equation}
    g\mu_bH_b^\mathrm{sat}=2S(4J_{\parallel}+J_{\bot}-D).
\end{equation}
At low $H_b$, the AFM state persists up to $H_b^\mathrm{f}$, where it transforms into the spin-flop state with a magnetization jump 
\begin{equation}
    \Delta M_b^\mathrm{f}= g \mu_B S \sqrt{D/(4J_{\parallel}+J_{\bot}-D)}.
\end{equation}
Further increase of $H_b$ leads to a gradual canting of the Ru moments, until the forced FM state is reached above $H_b^\mathrm{sat}$.

An analysis of the magnetic inelastic neutron scattering excitation spectra measured along the high-symmetry directions of the \CRTO \ lattice at \SI{5}{\kelvin} yields $J_{\parallel}=\SI{4.5}{\milli\eV}$, $J_{\bot}=\SI{3.0}{\milli\eV}$, and $D=\SI{4.5}{\milli\eV}$ \cite{KrautloherDraft2019,BertinshawDraft2021}. Although the applied Heisenberg model with these parameters, $S=1$, and $g=2$ reasonably reproduces the low-temperature, weak-field linear magnetization $M_a(H)$ -- yielding $\beta_a \approx 4.6\times10^{-3}\rm \mu_B/Ru\ T^{-1}$ vs. the measured $(5.75\pm0.01)\times10^{-3}\rm \mu_B/Ru\ T^{-1}$ -- it overestimates the critical fields by more than an order of magnitude: \SI{149}{\tesla}, \SI{285}{\tesla}, and \SI{440}{\tesla} compared with the experimentally extracted values of \SI{6}{\tesla}, \SI{10.5}{\tesla}, and \SI{20.9}{\tesla} for $H_b^\mathrm{f}$, $H_b^\mathrm{sat}$, and $H_a^\mathrm{sat}$, respectively. Such a significant discrepancy indicates a substantial divergence from pure Heisenberg physics in \CRTO. This is supported by the first-order phase boundaries between the PM and ordered AFM states. Unlike canonical AFMs, which exhibit continuous (second-order) PM–AFM transitions \cite{King1979}, \CRTO \ undergoes a first-order magnetoelastic phase transition with pronounced hysteresis and phase coexistence. This indicates the presence of strong elastic strain due to disproportionate changes in the lattice parameters at $T_\mathrm{N}$. Strong spin–phonon coupling can substantially weaken long‑range Néel order in \CRTO. Cooperative electron–electron and electron–lattice interactions with coupled structural and magnetic order parameters have also been brought into connection with the compound’s exceptional sensitivity to light \cite{Rabinovich2024}. In our model Hamiltonian (\cref{eq:Hamiltonian}) the strong spin-phonon coupling has not been incorporated. This can be accomplished by augmenting the bond-dependent exchange and by including explicit spin-phonon terms \cite{Aoyama2016,SS2025}, however this is beyond the scope of this experimental work. For the sake of completeness, we mention the prominent example of the manganites where also a strong renormalization of the electronic and magnetic properties has been reported. There it has been attributed to the combined effect of strong electron-phonon coupling and the double exchange mechanism between itinerant electrons and localized spins. Together, these interactions give for instance rise to colossal magnetoresistance and field-driven metal-insulator transitions of enormous magnitude \cite{tokura2006critical}. This warrants a closer look at the magnetotransport properties of \CRTO \ and a comparison with the rich magnetic phase diagram established above.

In the next section, we show that \CRTO \ remains insulating through the spin-flop transition at $H_b^{f}$ but becomes metallic in the field-induced, saturated FM state above $H_b^{sat}$. This insulator-to-metal transition indicates that electron itinerancy and low-energy electron–hole excitations -- features beyond a minimal Heisenberg spin-only model -- become also relevant near the field-driven phase boundary.

\subsection{Magnetoresistance properties}
\begin{figure}%[b]
    \includegraphics[width=\linewidth]{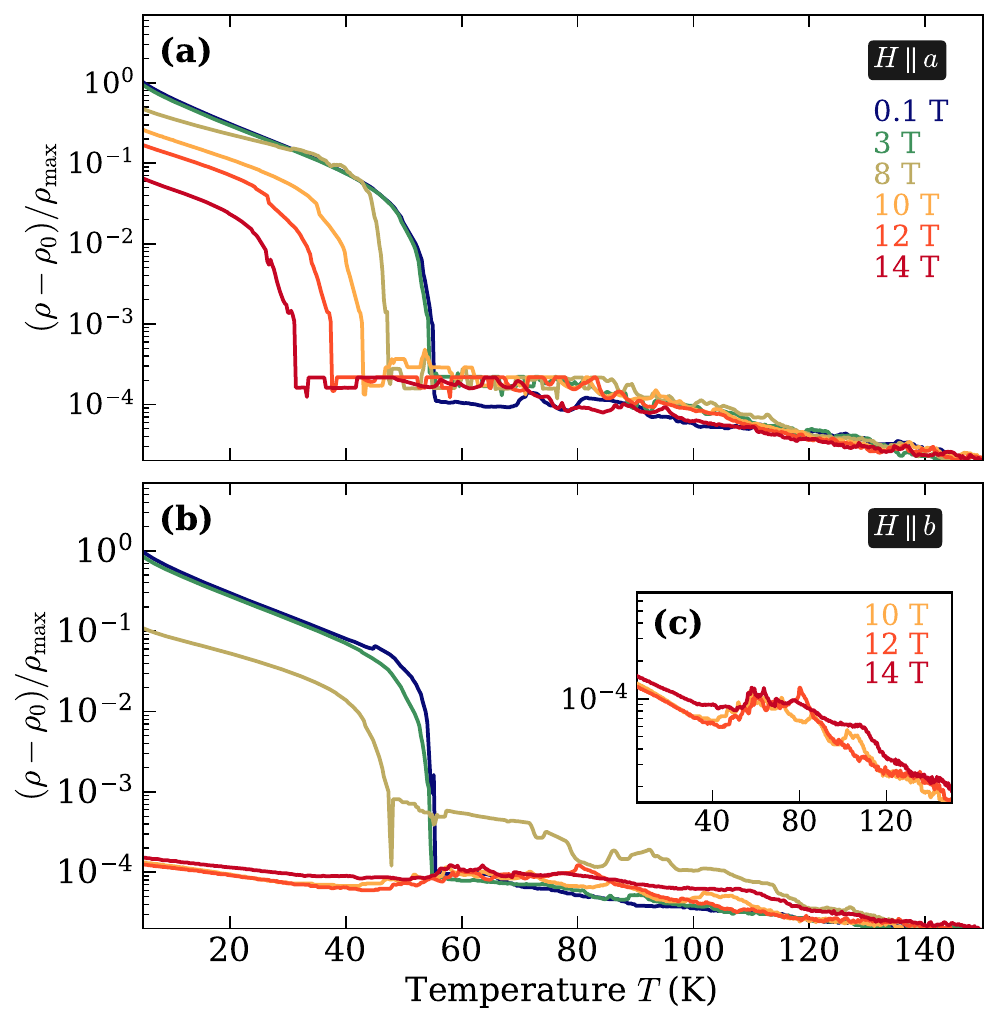}
    \phantomsubfloat{\label{fig:figure05_a}}
    \phantomsubfloat{\label{fig:figure05_b}}
    \phantomsubfloat{\label{fig:figure05_c}}
    \caption{\label{fig:5} Temperature dependence of the normalized in-plane, low-frequency (\SI{38}{\hertz}) resistivity, $(\rho-\rho_0)/\rho_\mathrm{max}$, of \CRTO \ measured for different magnetic fields. Fields up to \SI{14}{\tesla} are applied along the $a$-axis (a) and $b$-axis (b). (c) Magnified view for fields \SI{\geq10}{\tesla} along the $b$-axis.
    }
\end{figure}
%Peng2013,Gangshettiwar2020} 
In the absence of a magnetic field, Ti-substituted \CRTOx\ undergoes a sharp MIT at $T_\mathrm{N}$ upon cooling. The in-plane resistivity increases by three to four orders of magnitude \cite{Ke2011,Tsuda2013}. The field dependence of the MIT follows the $H$-$T$ magnetic phase diagram and gives rise to the colossal magnetoresistance observed at field-induced phase transitions \cite{Zhu2016}, though a reliable interpretation of the emerging phases requires minimal Ti substitution, as higher $x$ can lead to phase separation that complicates the analysis \cite{Peng2016a}.
Therefore, we focus here on \CRTO as such samples yield a simple $H$-$T$ phase diagram and show no signs of phase separation. Its Mott-insulating ground state 
is achieved by finely tuning the one-electron bandwidth $W$ through only 1\% substitution on the Ru site. The clean optical gap of \SI{0.45}{\eV} is sharply defined and entirely free of defect-induced excitations, as confirmed across the full spectral range down to \SI{1}{\milli\eV} \cite{Rabinovich2024}. The collapse of the Mott state above $T_N$ gives rise to a metallic Drude response with a plasma frequency as large as \SI{2.8}{\eV}, reflecting a substantial itinerant charge carrier density.

\Cref{fig:5} shows FC temperature dependencies of the normalized in-plane resistivity, $(\rho-\rho_0)/\rho_\mathrm{max}$, for different representative magnetic fields. The data set illustrates that an applied magnetic field lowers the temperature at which the metal insulator transition, signaled by a sharp resistivity increase, occurs. The temperature change of the MIT follows the quadratic relation of \cref{eq:TNvsH2} describing the field induced suppression of the Néel temperature. The steep MIT related resistivity upturn is present for all fields up to \SI{14}{\tesla} when the field is applied along the $a$-axis (\cref{fig:figure05_a}). When the field is aligned with the easy $b$-axis, the MIT transition follows the same trend for low magnetic fields, but it disappears 
for fields above \SI{10}{\tesla} (\cref{fig:figure05_b,fig:figure05_c}). 
Hence, \CRTO \ remains in the conductive PM phase at all temperatures when the field is applied along the easy $b$-axis and exceeds the critical field value $H_b^\mathrm{sat}$. In this high-field regime, the temperature-driven crossover from the field-forced FM state to the magnetically disordered state is accompanied by fluctuations in the resistivity. We attribute these to the formation and growth of ferromagnetic domains as domain walls are bound to alter the microscopic spatial current distribution.
\begin{figure}%[h!]
    \includegraphics[width=\linewidth]{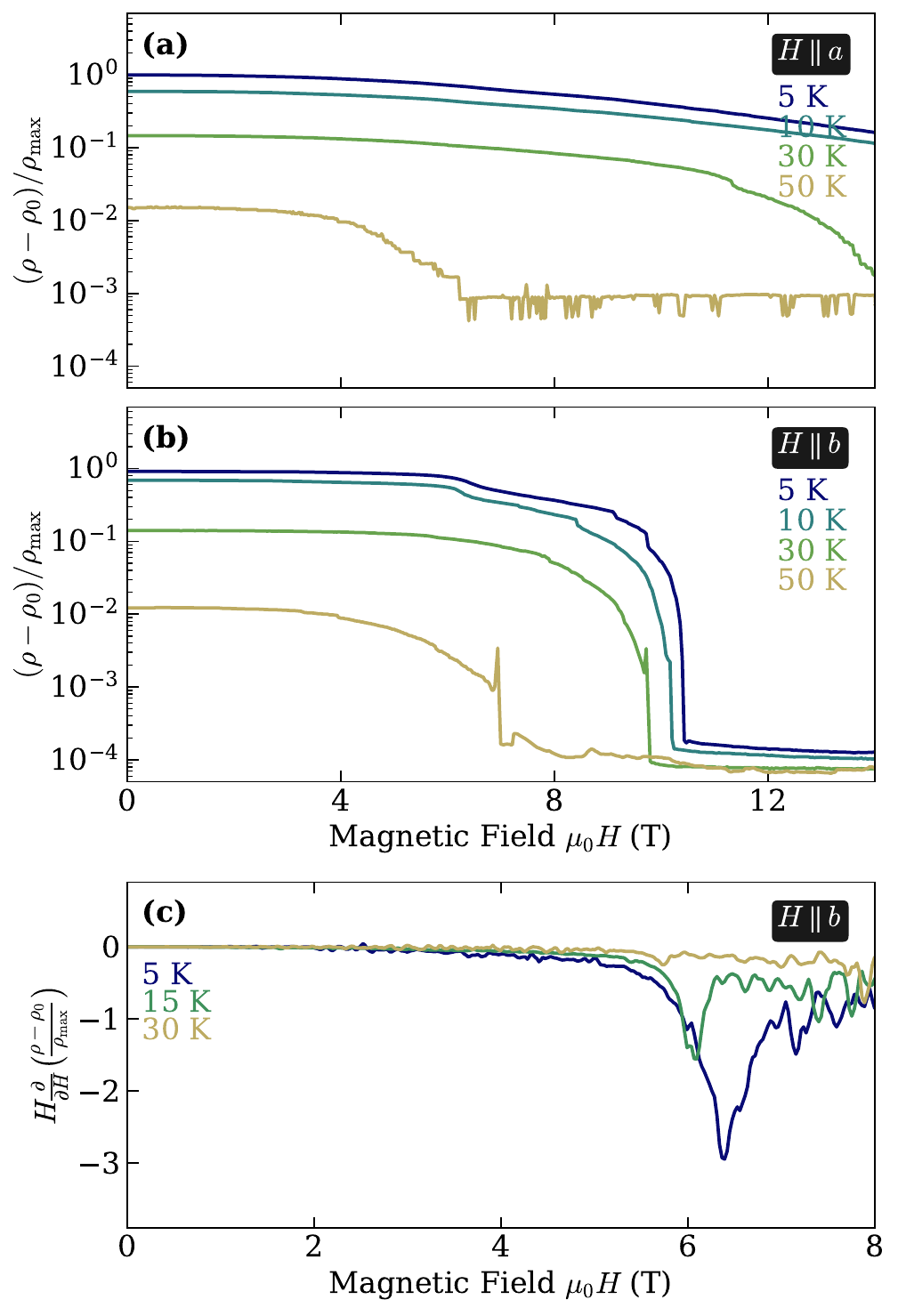}
    \phantomsubfloat{\label{fig:figure06_a}}
    \phantomsubfloat{\label{fig:figure06_b}}
    \phantomsubfloat{\label{fig:figure06_c}}
    \caption{\label{fig:6} Magnetic-field dependence of the normalized in-plane, low-frequency (\SI{38}{\hertz}) resistivity, $(\rho-\rho_0)/\rho_\mathrm{max}$, measured at temperatures below $T_\mathrm{N}$ with fields applied along the $a$-axis (a) and $b$-axis (b). (c) Field-weighted derivative, $H\partial((\rho-\rho_0)/\rho_\mathrm{max})/\partial H$, for the field applied along the $b$-axis. The minimum in the derivative corresponds to a slight drop in in-plane resistivity as a canted FM moment develops along the $b$-axis at the field-induced spin-flop transition above $H_b^\mathrm{f}$.
    }
\end{figure}

We also examined the isothermal field dependence of the in-plane resistivity to continue mapping the electronic MIT onto the $H$-$T$ magnetic phase diagram in \cref{fig:4}. \Cref{fig:6} shows ZFC curves of the normalized in-plane resistivity, $(\rho-\rho_0)/\rho_\mathrm{max}$, versus magnetic field at selected temperatures below $T_\mathrm{N}$. The negative magnetoresistance across the field-driven transition from the G-AFM to PM state closely tracks the field-dependent magnetization in \cref{fig:2}. For fields up to \SI{14}{\tesla} along the $a$-axis in \cref{fig:figure06_a}, the highly conductive regime appears only near $T_\mathrm{N}$. In contrast, for fields along the easy $b$-axis, the normalized resistivity in \cref{fig:figure06_b} drops by two to four orders of magnitude in the forced FM state above $H_b^\mathrm{sat}$, while the spin-flop phase for ($H_b^\mathrm{f} < H < H_b^\mathrm{sat}$) remains insulating. Across the spin-flop transition we find only small, yet well-reproducible, step-like resistivity drops, which correspond to minima in the derivative of the field-dependent normalized resistivity at $H_b^\mathrm{f}$ (\cref{fig:figure06_c}).

\begin{figure}%[h!]
    \includegraphics[width=\linewidth]{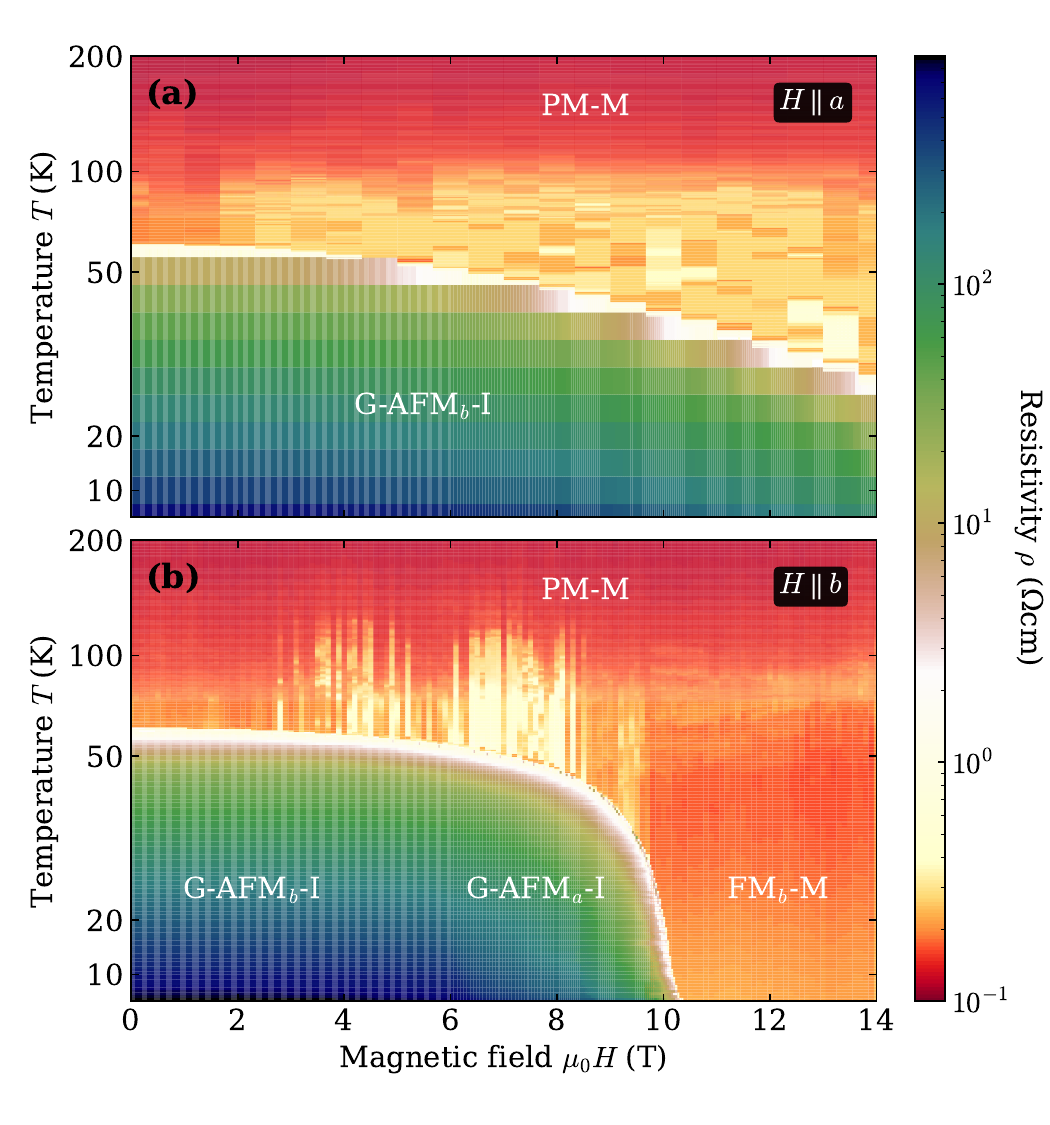}
    \phantomsubfloat{\label{fig:figure07_a}}
    \phantomsubfloat{\label{fig:figure07_b}}
    \caption{\label{fig:7} Color maps of the in-plane resistivity, $\rho$, as functions of temperature $T$ and magnetic field $H$, with $H$ applied along the $a$-axis \textbf{(a)} and $b$-axis \textbf{(b)}. Low-frequency (\SI{38}{\hertz}) resistivity data are stitched from measurement protocols optimized for the insulating (green) and metallic (orange) regions: in the insulating regime we use $\rho(H)$ field sweeps at fixed $T$; in the metallic regime we use $\rho(T)$ sweeps at fixed $H$, emphasizing phase-specific features.
    }
\end{figure}

In order to compile a more detailed phase diagram, the in-plane resistivity was measured as a function of both magnetic field and temperature at fine intervals and compiled into the detailed color map in \cref{fig:figure07_a,fig:figure07_b}. Sharply delineating the MIT, the resistivity color plot traces the phase boundary between the G-AFM and PM phases in the $H$-$T$ magnetic phase diagrams constructed from the $\chi(T)$ and $M(H)$ data in \cref{fig:figure04_a,fig:figure04_b}, respectively. Unlike the magnetization and magnetic susceptibility, the magnetoresistance is not a purely thermodynamic quantity and is sensitive to microscopic details and local features. Its behavior can for instance depend on the magnetic domain structure, the domain boundaries and defect related pinning. Therefore, the magnetoresistance can reveal complementary information that is absent in the magnetic data. In particular, in the metallic phase, as the temperature decreases at constant field, the resistivity increases and pronounced irregularities emerge upon approaching the MIT. Such electrical instabilities become most pronounced near the critical point around \SI{8}{\tesla} along the $b$-axis, where multiple phase boundary lines converge. These instabilities can be associated with short-range AFM correlations persisting above $T_N$ in \CRTO\ across the broad $T$-$H$ phase region near the critical point.
 
Another notable feature appears in \cref{fig:figure07_b} for fields above the critical $H_b^\mathrm{sat}$. A distinct boundary on the resistivity color plot marks the crossover between the field-forced FM state and the high-temperature PM behavior restored by thermal fluctuations, coinciding with the line traced by the star symbols in \cref{fig:figure04_b}. %). [THIS CAN BE SEEN BUT NOT THAT THE FLUCTUATIONS ARE MORE PRONOUNCED NEAR 8T]
As illustrated in \cref{fig:figure05_c}, this phase region also exhibits electrical instabilities, here associated with the dynamics of FM domains. Counterintuitively, the saturated FM state is more resistive than the high-temperature magnetically disordered state, consistent with the overall decrease of resistivity with increasing temperature seen in \cref{fig:figure05_c}. A similar trend is observed in pristine \CROb, where the field-forced FM state shows higher resistivity than the PM state, attributed to the suppression of interbilayer hopping governed by orbital degrees of freedom in the fully polarized FM configuration \cite{Lin2005}. This effect stems from the bilayer ruthenate structure, where spin–orbit coupling causes FM alignment to hinder electron hopping more effectively than PM disorder.

In contrast to the significant resistance change observed for the crossover between the field-forced FM state and PM state restored by the thermal fluctuations at higher temperature, only a slight drop in resistivity is observed across the spin-flop transition (\cref{fig:figure06_b,fig:figure06_c}), where a canted FM moment emerges above $H_b^\mathrm{f}$ (\cref{fig:figure03_b}). The net FM component along the $b$-axis partially aligns neighboring Ru spins across bilayers, 
reducing spin scattering. Virtual interlayer hopping enhances Ru $4d$ electron itinerancy via a double-exchange-like mechanism, without being subject to the orbital‑selection rules that suppress hopping in the fully polarized FM state. The normalized resistivity decreases by less than 5\%, a change barely discernible in the resistivity colormap (green region, \cref{fig:figure07_b}). Both the G-AFM-$b$ ground state and the spin-flop G-AFM-$a$ state remain Mott insulating. The extreme renormalization of the critical field $H_b^\mathrm{f}$ relative to the prediction of the minimal Heisenberg spin-only model (\cref{eq:Hamiltonian}) is therefore primarily attributed to strong spin–phonon coupling. Unlike the G-AFM-I to PM-M transition, the role of electron itinerancy and low-energy electron–hole excitations remains negligible.

\section{Conclusions}
We have systematically investigated the magnetic and electronic phase transitions in bilayer perovskite \CRTO \ in the presence of in-plane magnetic fields up to \SI{14}{\tesla}. Using anomalies in $\chi(T)$ under various fields, isothermal $M(H)$ traces below $T_\mathrm{N} = \SI{55}{\kelvin}$, as well as temperature and field dependent resistivity data, we have constructed the $H$-$T$ phase diagram of this compound. With the field along the easy $b$-axis, \CRTO \ exhibits a first-order spin-flop transition at $H_b^\mathrm{f}\approx\SI{6}{\tesla}$, while retaining its Mott insulating state across the transition. Subsequently, a transition to a forced FM metallic phase occurs above $H_b^\mathrm{sat}\approx\SI{10.5}{\tesla}$. When the field is applied along the hard $a$-axis, neither the spin-flop nor the forced-FM phase emerges below $\SI{14}{\tesla}$. In the presence of strong electron-lattice coupling, \CRTO \ undergoes the first-order magnetoelastic G-AFM-I to PM-M transitions, yet it exhibits the magnetic $H$-$T$ phase diagram characteristic of canonical anisotropic AFMs. The critical fields $H_b^{f}$ and $H_b^{sat}$ are hence suppresses by more than an order of magnitude beyond the predictions of a minimal Heisenberg spin-only model, that accounts for the exchange interactions in \CRTO. We anticipate that these results will stimulate further theoretical studies to quantitatively describe the magnetic phase diagram of Mott AFMs in the presence of strong electron-lattice coupling. On the experimental side, direct probes are necessary to detail the impact of magnetic fields on structural distortions in \CRTO \ and precisely quantify magnetoelastic effects at the phase boundaries.

\begin{acknowledgments}
We gratefully acknowledge A. N. Yaresko, H. Gretarsson, A. Kirilyuk, G. Khaliullin, L. Wang for fruitful discussions. This work was supported by the European Research Council under Advanced Grant No. 101141844 (SpecTera).
\end{acknowledgments}

\bibliography{CRTOBibMay2026}

@Article{Peng2016b,
  author       = {Peng, Jin and Liu, J. Y. and Hu, J. and Mao, Z. Q. and Zhang, F. M. and Wu, X. S.},
  date         = {2016-01},
  journal = {Scientific Reports},
  title        = {Magnetic phase separation in double layer ruthenates {Ca$_3$(Ru$_{1−x}$Ti$_x$)$_2$O$_7$}},
  doi          = {10.1038/srep19462},
  issn         = {2045-2322},
  number       = {1},
  volume       = {6},
  publisher    = {Springer Science and Business Media LLC},
}

@Article{Cao1999,
  author       = {Cao, G and Alexander, C.S and McCall, S and Crow, J.E and Guertin, R.P},
  year         = {1999},
  journal = {Materials Science and Engineering: B},
  title        = {From antiferromagnetic insulator to ferromagnetic metal: a brief review of the layered ruthenates},
  doi          = {10.1016/s0921-5107(99)00055-0},
  issn         = {0921-5107},
  number       = {1–2},
  pages        = {76--82},
  volume       = {63},
  publisher    = {Elsevier BV},
}

@Article{Bertinshaw2019,
  author       = {Bertinshaw, J. and Gurung, N. and Jorba, P. and Liu, H. and Schmid, M. and Mantadakis, D. T. and Daghofer, M. and Krautloher, M. and Jain, A. and Ryu, G. H. and Fabelo, O. and Hansmann, P. and Khaliullin, G. and Pfleiderer, C. and Keimer, B. and Kim, B.J.},
  year         = {2019},
  journal = {Phys. Rev. Lett.},
  title        = {Unique Crystal Structure of $\mathrm{Ca}_{2}\mathrm{Ru}\mathrm{O}_{4}$ in the Current Stabilized Semimetallic State},
  doi          = {10.1103/physrevlett.123.137204},
  issn         = {1079-7114},
  number       = {13},
  pages        = {137204},
  volume       = {123},
  publisher    = {American Physical Society (APS)},
}

@Article{Suen2024,
  author       = {Suen, C. T. and Marković, I. and Zonno, M. and Heinsdorf, N. and Zhdanovich, S. and Jo, N. H. and Schmid, M. and Hansmann, P. and Puphal, P. and Fürsich, K. and Smit, S. and Au-Yeung, C. and Zimmermann, V. and Zwartsenberg, B. and Krautloher, M. and Elfimov, I. S. and Koch, R. and Gorovikov, S. and Jozwiak, C. and Bostwick, A. and Franz, M. and Rotenberg, Eli and Keimer, B. and Damascelli, A.},
  year         = {2024},
  journal = {Nat. Phys.},
  title        = {Electronic response of a $\mathrm{M}$ott insulator at a current-induced insulator-to-metal transition},
  doi          = {10.1038/s41567-024-02629-3},
  issn         = {1745-2481},
  number       = {11},
  pages        = {1757--1763},
  volume       = {20},
  publisher    = {Springer Science and Business Media LLC},
}

@Article{Zhang2019,
  author       = {Zhang, Jiawei and McLeod, Alexander S. and Han, Qiang and Chen, Xinzhong and Bechtel, Hans A. and Yao, Ziheng and Gilbert Corder, S. N. and Ciavatti, Thomas and Tao, Tiger H. and Aronson, Meigan and Carr, G. L. and Martin, Michael C. and Sow, Chanchal and Yonezawa, Shingo and Nakamura, Fumihiko and Terasaki, Ichiro and Basov, D. N. and Millis, Andrew J. and Maeno, Yoshiteru and Liu, Mengkun},
  year         = {2019},
  journal = { Phys. Rev. X},
  title        = {Nano-Resolved Current-Induced Insulator-Metal Transition in the $\mathrm{M}$ott Insulator $\mathrm{Ca}_{2}\mathrm{Ru}\mathrm{O}_{4}$},
  doi          = {10.1103/physrevx.9.011032},
  issn         = {2160-3308},
  number       = {1},
  pages        = {011032},
  volume       = {9},
  publisher    = {American Physical Society (APS)},
}

@Article{Ikeda2000,
  author       = {Ikeda, Shin-Ichi and Maeno, Yoshiteru and Nakatsuji, Satoru and Kosaka, Masashi and Uwatoko, Yoshiya},
  year         = {2000},
  journal = {Phys. Rev. B},
  title        = {Ground state in $\mathrm{Sr}_{3}\mathrm{Ru}_{2}\mathrm{O}_{7}$: Fermi liquid close to a ferromagnetic instability},
  doi          = {10.1103/physrevb.62.r6089},
  issn         = {1095-3795},
  number       = {10},
  pages        = {R6089--R6092},
  volume       = {62},
  publisher    = {American Physical Society (APS)},
}

@Article{Snow2002,
  author       = {Snow, C. S. and Cooper, S. L. and Cao, G. and Crow, J. E. and Fukazawa, H. and Nakatsuji, S. and Maeno, Y.},
  year         = {2002},
  journal = {Phys. Rev. Lett.},
  title        = {Pressure-Tuned Collapse of the $\mathrm{M}$ott-Like State in $\mathrm{Ca}_{n+1}\mathrm{Ru}_{n}\mathrm{O}_{3n+1}$ (n=1,2): Raman Spectroscopic Studies},
  doi          = {10.1103/physrevlett.89.226401},
  issn         = {1079-7114},
  number       = {22},
  pages        = {226401},
  volume       = {89},
  publisher    = {American Physical Society (APS)},
}

@Article{Lin2005,
  author       = {Lin, X. N. and Zhou, Z. X. and Durairaj, V. and Schlottmann, P. and Cao, G.},
  year         = {2005},
  journal = {Phys. Rev. Lett.},
  title        = {Colossal Magnetoresistance by Avoiding a Ferromagnetic State in the $\mathrm{M}$ott System $\mathrm{Ca}_{3}\mathrm{Ru}_{2}\mathrm{O}_{7}$},
  doi          = {10.1103/physrevlett.95.017203},
  issn         = {1079-7114},
  number       = {1},
  pages        = {017203},
  volume       = {95},
  publisher    = {American Physical Society (APS)},
}

@Article{Maeno2001,
  author       = {Maeno, Yoshiteru and Rice, T. Maurice and Sigrist, Manfred},
  year         = {2001},
  journal = {Phys. Today},
  title        = {The Intriguing Superconductivity of Strontium Ruthenate},
  doi          = {10.1063/1.1349611},
  issn         = {1945-0699},
  number       = {1},
  pages        = {42--47},
  volume       = {54},
  publisher    = {AIP Publishing},
}

@Article{Antropov2021,
  author       = {Antropov, N. O. and Kravtsov, E. A. and Makarova, M. V. and Proglyado, V. V. and Keller, T. and Subbotin, I. A. and Pashaev, E. M. and Prutskov, G. V. and Vasiliev, A. L. and Chesnokov, Yu. M. and Bebenin, N. G. and Milyaev, M. A. and Ustinov, V. V. and Keimer, B. and Khaydukov, Yu. N.},
  year         = {2021},
  journal = {Phys. Rev. B},
  title        = {Tunable spin-flop transition in artificial ferrimagnets},
  doi          = {10.1103/physrevb.104.054414},
  issn         = {2469-9969},
  number       = {5},
  pages        = {054414},
  volume       = {104},
  publisher    = {American Physical Society (APS)},
}

@Article{Markovic2020,
  author       = {Marković, Igor and Watson, Matthew D. and Clark, Oliver J. and Mazzola, Federico and Abarca Morales, Edgar and Hooley, Chris A. and Rosner, Helge and Polley, Craig M. and Balasubramanian, Thiagarajan and Mukherjee, Saumya and Kikugawa, Naoki and Sokolov, Dmitry A. and Mackenzie, Andrew P. and King, Phil D. C.},
  year         = {2020},
  journal = {Proceedings of the National Academy of Sciences},
  title        = {Electronically driven spin-reorientation transition of the correlated polar metal $\mathrm{Ca}_{3}\mathrm{Ru}_{2}\mathrm{O}_{7}$},
  doi          = {10.1073/pnas.2003671117},
  issn         = {1091-6490},
  number       = {27},
  pages        = {15524--15529},
  volume       = {117},
  publisher    = {Proceedings of the National Academy of Sciences},
}

@Article{Rabinovich2024,
  author       = {Rabinovich, Ksenia S. and Yaresko, Alexander N. and Dawson, Robert D. and Krautloher, Maximilian J. and Priessnitz, Tim and Mathis, Yves‐Laurent and Kirilyuk, Andrei and Keimer, Bernhard and Boris, Alexander V.},
  year         = {2025},
  journal = {Advanced Functional Materials},
  title        = {Optically Driven Dynamics of a $\mathrm{M}$ott Insulator‐to‐Metal Transition},
  doi          = {10.1002/adfm.202416597},
  issn         = {1616-3028},
  number       = {10},
  volume       = {35},
 pages        = {2416597},
  publisher    = {Wiley},
}

@Article{Shapira1970,
  author       = {Shapira, Y. and Foner, S.},
  year         = {1970},
  journal = {Phys. Rev. B},
  title        = {Magnetic Phase Diagram of $\mathrm{MnF}_{2}$ from Ultrasonic and Differential Magnetization Measurements},
  doi          = {10.1103/physrevb.1.3083},
  issn         = {0556-2805},
  number       = {7},
  pages        = {3083--3096},
  volume       = {1},
  publisher    = {American Physical Society (APS)},
}

@Article{Shrestha2025,
  author       = {Shrestha, Sujan and Souri, Maryam and Dietl, Christopher J. and Pärschke, Ekaterina M. and Krautloher, Maximilian and Calderon Ortiz, Gabriel A. and Minola, Matteo and Shi, Xiatong and Boris, Alexander V. and Hwang, Jinwoo and Khaliullin, Giniyat and Cao, Gang and Keimer, Bernhard and Kim, Jong-Woo and Kim, Jungho and Seo, Ambrose},
  year         = {2025},
  journal = {Nat. Commun.},
  title        = {Tunable magnons of an antiferromagnetic $\mathrm{M}$ott insulator via interfacial metal-insulator transitions},
  doi          = {10.1038/s41467-025-58922-z},
  issn         = {2041-1723},
  number       = {1},
  volume       = {16},
  pages     = {3592},
  publisher    = {Springer Science and Business Media LLC},
}

@Article{Cao1997,
  author    = {Cao, G. and McCall, S. and Crow, J. E. and Guertin, R. P.},
  title     = {Observation of a Metallic Antiferromagnetic Phase and Metal to Nonmetal Transition in {${\mathrm{Ca}}_{3}{\mathrm{Ru}}_{2}{\mathrm{O}}_{7}$}},
  doi       = {10.1103/PhysRevLett.78.1751},
  issue     = {9},
  pages     = {1751--1754},
  url       = {https://link.aps.org/doi/10.1103/PhysRevLett.78.1751},
  volume    = {78},
  journal   = {Phys. Rev. Lett.},
  month     = {Mar},
  numpages  = {0},
  publisher = {American Physical Society},
  year      = {1997},
}

@Article{Bao2008,
  author    = {Bao, Wei and Mao, Z. Q. and Qu, Z. and Lynn, J. W.},
  title     = {Spin Valve Effect and Magnetoresistivity in Single Crystalline {${\mathrm{Ca}}_{3}{\mathrm{Ru}}_{2}{\mathrm{O}}_{7}$}},
  doi       = {10.1103/PhysRevLett.100.247203},
  issue     = {24},
  pages     = {247203},
  url       = {https://link.aps.org/doi/10.1103/PhysRevLett.100.247203},
  volume    = {100},
  journal   = {Phys. Rev. Lett.},
  month     = {Jun},
  numpages  = {4},
  publisher = {American Physical Society},
  year      = {2008},
}

@Article{Sokolov2019,
  author    = {Sokolov, DA and Kikugawa, N and Helm, T and Borrmann, H and Burkhardt, U and Cubitt, R and White, JS and Ressouche, E and Bleuel, M and Kummer, K and others},
  title     = {Metamagnetic texture in a polar antiferromagnet},
  number    = {7},
  pages     = {671--677},
  url       = {https://www.nature.com/articles/s41567-019-0501-0},
  volume    = {15},
  journal   = {Nat. Phys.},
  publisher = {Nature Publishing Group},
  year      = {2019},
}

@Article{Ke2011,
  author    = {Ke, X. and Peng, J. and Singh, D. J. and Hong, Tao and Tian, Wei and Dela Cruz, C. R. and Mao, Z. Q.},
  title     = {Emergent electronic and magnetic state in {${\mathrm{Ca}}_{3}{\mathrm{Ru}}_{2}{\mathrm{O}}_{7}$} induced by {Ti} doping},
  doi       = {10.1103/PhysRevB.84.201102},
  issue     = {20},
  pages     = {201102},
  url       = {https://link.aps.org/doi/10.1103/PhysRevB.84.201102},
  volume    = {84},
  journal   = {Phys. Rev. B},
  month     = {Nov},
  numpages  = {4},
  publisher = {American Physical Society},
  year      = {2011},
}

@Article{Tsuda2013,
  author    = {S. Tsuda and N. Kikugawa and K. Sugii and S. Uji and S. Ueda and M. Nishio and Y. Maeno},
  title     = {Mott transition extremely sensitive to impurities in {${\mathrm{Ca}}_{3}{\mathrm{Ru}}_{2}{\mathrm{O}}_{7}$} revealed by hard {X}-ray photoemission studies},
  doi       = {10.1103/PhysRevB.87.241107},
  number    = {24},
  pages     = {241107},
  volume    = {87},
  journal   = {Phys. Rev. B},
  month     = {jun},
  publisher = {American Physical Society ({APS})},
  year      = {2013},
}

@Article{Gomonay2018,
  author       = {Gomonay, O. and Baltz, V. and Brataas, A. and Tserkovnyak, Y.},
  year         = {2018},
  journal = {Nat. Phys.},
  title        = {Antiferromagnetic spin textures and dynamics},
  doi          = {10.1038/s41567-018-0049-4},
  issn         = {1745-2481},
  number       = {3},
  pages        = {213--216},
  volume       = {14},
  publisher    = {Springer Science and Business Media LLC},
}

@Article{Baltz2018,
  author       = {Baltz, V. and Manchon, A. and Tsoi, M. and Moriyama, T. and Ono, T. and Tserkovnyak, Y.},
  year         = {2018},
  journal = {Reviews of Modern Physics},
  title        = {Antiferromagnetic spintronics},
  doi          = {10.1103/revmodphys.90.015005},
  issn         = {1539-0756},
  number       = {1},
  pages        = {015005},
  volume       = {90},
  publisher    = {American Physical Society (APS)},
}

@Article{Chen2024,
  author       = {Chen, Hongyu and Liu, Li and Zhou, Xiaorong and Meng, Ziang and Wang, Xiaoning and Duan, Zhiyuan and Zhao, Guojian and Yan, Han and Qin, Peixin and Liu, Zhiqi},
  year         = {2024},
  journal = {Adv. Mater.},
  title        = {Emerging Antiferromagnets for Spintronics},
  doi          = {10.1002/adma.202310379},
  issn         = {1521-4095},
  number       = {14},
  volume       = {36},
 pages        = {2310379},
  publisher    = {Wiley},
}

@Article{Faure2023,
  author       = {Faure, Q. and Dashwood, C. D. and Colin, C. V. and Johnson, R. D. and Ressouche, E. and Stenning, G. B. G. and Spratt, J. and McMorrow, D. F. and Perry, R. S.},
  year         = {2023},
  journal = {Physical Review Research},
  title        = {Magnetic structure and field dependence of the cycloid phase mediating the spin reorientation transition in $\mathrm{Ca}_{3}\mathrm{Ru}_{2}\mathrm{O}_{7}$},
  doi          = {10.1103/physrevresearch.5.013040},
  issn         = {2643-1564},
  number       = {1},
  pages        = {013040},
  volume       = {5},
  publisher    = {American Physical Society (APS)},
}

@Article{Bohnenbuck2008,
  author       = {Bohnenbuck, B. and Zegkinoglou, I. and Strempfer, J. and Schüßler-Langeheine, C. and Nelson, C. S. and Leininger, Ph. and Wu, H.-H. and Schierle, E. and Lang, J. C. and Srajer, G. and Ikeda, S. I. and Yoshida, Y. and Iwata, K. and Katano, S. and Kikugawa, N. and Keimer, B.},
  year         = {2008},
  journal = {Phys. Rev. B},
  title        = {Magnetic structure and orbital state of $\mathrm{Ca}_{3}\mathrm{Ru}_{2}\mathrm{O}_{7}$ investigated by resonant x-ray diffraction},
  doi          = {10.1103/physrevb.77.224412},
  issn         = {1550-235X},
  number       = {22},
  pages        = {224412},
  volume       = {77},
  publisher    = {American Physical Society (APS)},
}

@book{Krautloher2018,
  author    = {Krautloher, Maximilian},
  title     = {Neutron scattering studies on layered ruthenates},
  doi       = {10.18419/opus-10298},
    url       = {https://elib.uni-stuttgart.de/handle/11682/10315},
  copyright = {info:eu-repo/semantics/openAccess},
  publisher = {Universität Stuttgart},
  year      = {2018},
}

@Article{Peng2016a,
  author       = {Peng, Jin and Liu, J Y and Gu, Xiaomin and Zhou, Guotai and Wang, Wei and Hu, J and Zhang, F M and Wu, X S},
  year         = {2016},
  journal = {Journal of Physics D: Applied Physics},
  title        = {Extremely large anisotropic transport caused by electronic phase separation in $\mathrm{T}$i-doped $\mathrm{Ca}_{3}\mathrm{Ru}_{2}\mathrm{O}_{7}$},
  doi          = {10.1088/0022-3727/49/24/245004},
  issn         = {1361-6463},
  number       = {24},
  pages        = {245004},
  volume       = {49},
  publisher    = {IOP Publishing},
}

@article{Zhu2016,
  title = {Colossal Magnetoresistance in a Mott Insulator via Magnetic Field-Driven Insulator-Metal Transition},
  author = {Zhu, M. and Peng, J. and Zou, T. and Prokes, K. and Mahanti, S. D. and Hong, T. and Mao, Z. Q. and Liu, G. Q. and Ke, X.},
  journal = {Phys. Rev. Lett.},
  volume = {116},
  issue = {21},
  pages = {216401},
  numpages = {5},
  year = {2016},
  month = {May},
  publisher = {American Physical Society},
  doi = {10.1103/PhysRevLett.116.216401},
  url = {https://link.aps.org/doi/10.1103/PhysRevLett.116.216401}
}

@Article{Heller1966,
  author       = {Heller, Peter},
  year         = {1966},
  journal = {Physical Review},
  title        = {Nuclear-Magnetic-Resonance Studies of Critical Phenomena in {MnF$_2$}. {I.} {T}ime-Average Properties},
  doi          = {10.1103/physrev.146.403},
  issn         = {0031-899X},
  number       = {2},
  pages        = {403},
  volume       = {146},
  publisher    = {American Physical Society (APS)},
}

@Unpublished{KrautloherDraft2019,
author    = {M. J. Krautloher and J. Bertinshaw and J. Porras and D. G. Joshi and H. Trepka and M. Hepting and D. Adroja and H. Walker and A. Ivanov and M. Enderle and U. Stuhr and B. J. Kim and B. Keimer},
year         = {Unpublished,2019},
  title  = {Impurity induced localization of low-energy spin excitations in bilayer ruthenate {Ca$_3$Ru$_2$O$_7$}},
}

@Unpublished{BertinshawDraft2021,
  author       = {J. Bertinshaw and H. Suzuki and H. Takahashi and M. Krautloher and K. S. Rabinovich and G. Khaliullin and A. V. Boris and H. Gretarsson and B. Keimer},
year         = {Unpublished,2021},
  title  = {The emergence of spin-orbit excitations in {Ti}-doped {Ca$_3$Ru$_2$O$_7$}},
}

@Unpublished{BertinshawNSDraft2020,
  author       = {J. Bertinshaw and M. L. Krautloher and L. Wang and H. Trepka and J. Porras and M. Hepting and S. Francoual and E. Ressouche and B. J. Kim and B. Keimer},
year         = {Unpublished,2020},
  title  = {Extreme sensitivity to perturbation in {Ti}-doped {Ca$_3$Ru$_2$O$_7$}},
}

@Article{King1979,
  author       = {King, A. R. and Rohrer, H.},
  year         = {1979},
  journal = {Physical Review B},
  title        = {Spin-flop bicritical point in {MnF$_2$}},
  doi          = {10.1103/physrevb.19.5864},
  issn         = {0163-1829},
  number       = {11},
  pages        = {5864},
  volume       = {19},
  publisher    = {American Physical Society (APS)},
}

@Article{Aoyama2016,
  author    = {Aoyama, Kazushi and Kawamura, Hikaru},
  title     = {Spin-Lattice-Coupled Order in {Heisenberg} Antiferromagnets on the Pyrochlore Lattice},
  doi       = {10.1103/PhysRevLett.116.257201},
  issue     = {25},
  pages     = {257201},
  url       = {https://link.aps.org/doi/10.1103/PhysRevLett.116.257201},
  volume    = {116},
  journal   = {Phys. Rev. Lett.},
  month     = {Jun},
  numpages  = {5},
  publisher = {American Physical Society},
  year      = {2016},
}

@Article{SS2025,
  author    = {SS, Jayakrishnan and Bansal, Dipanshu},
  title     = {Effect of spin-phonon coupling on phonons and magnons in the antiferromagnet {NiO}},
  doi       = {10.1103/PhysRevB.111.104306},
  issue     = {10},
  pages     = {104306},
  url       = {https://link.aps.org/doi/10.1103/PhysRevB.111.104306},
  volume    = {111},
  journal   = {Phys. Rev. B},
  month     = {Mar},
  numpages  = {12},
  publisher = {American Physical Society},
  year      = {2025},
}

@article{nakamura2013electric,
  title={Electric-field-induced metal maintained by current of the Mott insulator {Ca$_2$RuO$_4$}},
  author={Nakamura, Fumihiko and Sakaki, Mariko and Yamanaka, Yuya and Tamaru, Sho and Suzuki, Takashi and Maeno, Yoshiteru},
  journal={Scientific reports},
  volume={3},
  number={1},
  pages={2536},
  year={2013},
  publisher={Nature Publishing Group UK London}
}

@article{okazaki2013current,
  title={Current-induced gap suppression in the Mott insulator {Ca$_2$RuO$_4$}},
  author={Okazaki, Ryuji and Nishina, Yasuo and Yasui, Yukio and Nakamura, Fumihiko and Suzuki, Takashi and Terasaki, Ichiro},
  journal={Journal of the Physical Society of Japan},
  volume={82},
  number={10},
  pages={103702},
  year={2013},
  publisher={The Physical Society of Japan}
}

@article{leon2025hybrid,
  title={Hybrid d/p-wave altermagnetism in {Ca$_3$Ru$_2$O$_7$} and strain-controlled spin splitting},
  author={Le{\'o}n, Andrea and Autieri, Carmine and Brumme, Thomas and Gonz{\'a}lez, Jhon W},
  journal={npj Quantum Materials},
  volume={10},
  number={1},
  pages={98},
  year={2025},
  publisher={Nature Publishing Group UK London}
}

@article{cuono2023orbital,
  title={Orbital-selective altermagnetism and correlation-enhanced spin-splitting in strongly-correlated transition metal oxides},
  author={Cuono, Giuseppe and Sattigeri, Raghottam M and Skolimowski, Jan and Autieri, Carmine},
  journal={Journal of Magnetism and Magnetic Materials},
  volume={586},
  pages={171163},
  year={2023},
  publisher={Elsevier}
}

@article{fisher1962relation,
  title={Relation between the specific heat and susceptibility of an antiferromagnet},
  author={Fisher, Michael E},
  journal={Philosophical Magazine},
  volume={7},
  number={82},
  pages={1731--1743},
  year={1962},
  publisher={Taylor \& Francis}
}

@article{tokura2006critical,
  title={Critical features of colossal magnetoresistive manganites},
  author={Tokura, Y},
  journal={Reports on Progress in Physics},
  volume={69},
  number={3},
  pages={797--851},
  year={2006}
}

\end{document}